\documentclass[12pt,preprint]{aastex}

\usepackage{graphicx}
\usepackage[space]{grffile}
\usepackage{latexsym}
\usepackage{amsfonts,amsmath,amssymb}
\usepackage{url}
\usepackage[utf8]{inputenc}
\usepackage{fancyref}
\usepackage{hyperref}
\hypersetup{colorlinks=false,pdfborder={0 0 0},}
\def\rearth{\mbox{\,R}_\oplus}
\def\mearth{\mbox{\,M}_\oplus}
\usepackage[usenames,dvipsnames]{color}

\newcommand\gcmc{\,\mbox{g cm}^{-3}}

\newcommand{\ikt}{{\it Kepler}}
\newcommand{\ik}{{\it Kepler~}}
\title{Advances in Exoplanet Science from \ik}

\author{Jack J. Lissauer\footnote{NASA Ames Research Center, Moffett Field, CA
  94035, USA}, Rebekah I. Dawson\footnote{Department of Astronomy,
  University of California, Berkeley, CA 94720, USA},~ \& Scott
Tremaine\footnote{Institute for Advanced Study, Princeton, NJ 08540, USA}}

\slugcomment{Nature, submitted}

\begin{document}

\maketitle

\noindent {\bf Numerous telescopes and techniques have been used to find and study extrasolar planets, but none has been more successful than NASA's \ik\ Space Telescope. \ik has discovered the majority of known exoplanets, the smallest planets to orbit normal stars, and the worlds most likely to be similar to our home planet. Most importantly, \ik has provided our first look at  typical characteristics of planets and planetary systems for planets with sizes as small as and orbits as large as those of the Earth.}

\bigskip

\ik is a 0.95 m aperture space telescope launched by NASA in 2009 \citep{bor10,koch10}. \ik identifies those exoplanets whose orbits happen to appear edge-on by searching for  periodic dips caused by planetary transits (partial eclipses) of the stellar discs. Above Earth's atmosphere, and in an Earth-trailing heliocentric orbit away from the glare and thermal variations of low Earth orbit, \ik monitored the brightness of more than $10^5$ stars at 30-minute cadence for four years. \ikt's unique asset is an unprecedented photometric precision of $\sim 30$ parts per million (ppm) for 12th magnitude stars with data binned in 6.5 hour intervals \citep{gil11}. This time interval is used as a benchmark because the Earth takes 13 hours to transit the Sun as viewed by a distant observer in the ecliptic plane, and observers slightly away from the ecliptic view a transit of shorter duration. Such high-precision measurements are only possible in space, where stars do not twinkle, and are required to search for Earth-like worlds because the transit of such a planet across the disc of a Sun-like star blocks only 80 ppm of the stellar flux. For comparison, the transit of a Jupiter-size planet across a similar star blocks 1\% of the flux, and this dip is straightforward to detect using ground-based telescopes.

\begin{figure}[h!]
\begin{center}
\includegraphics[width=\columnwidth]{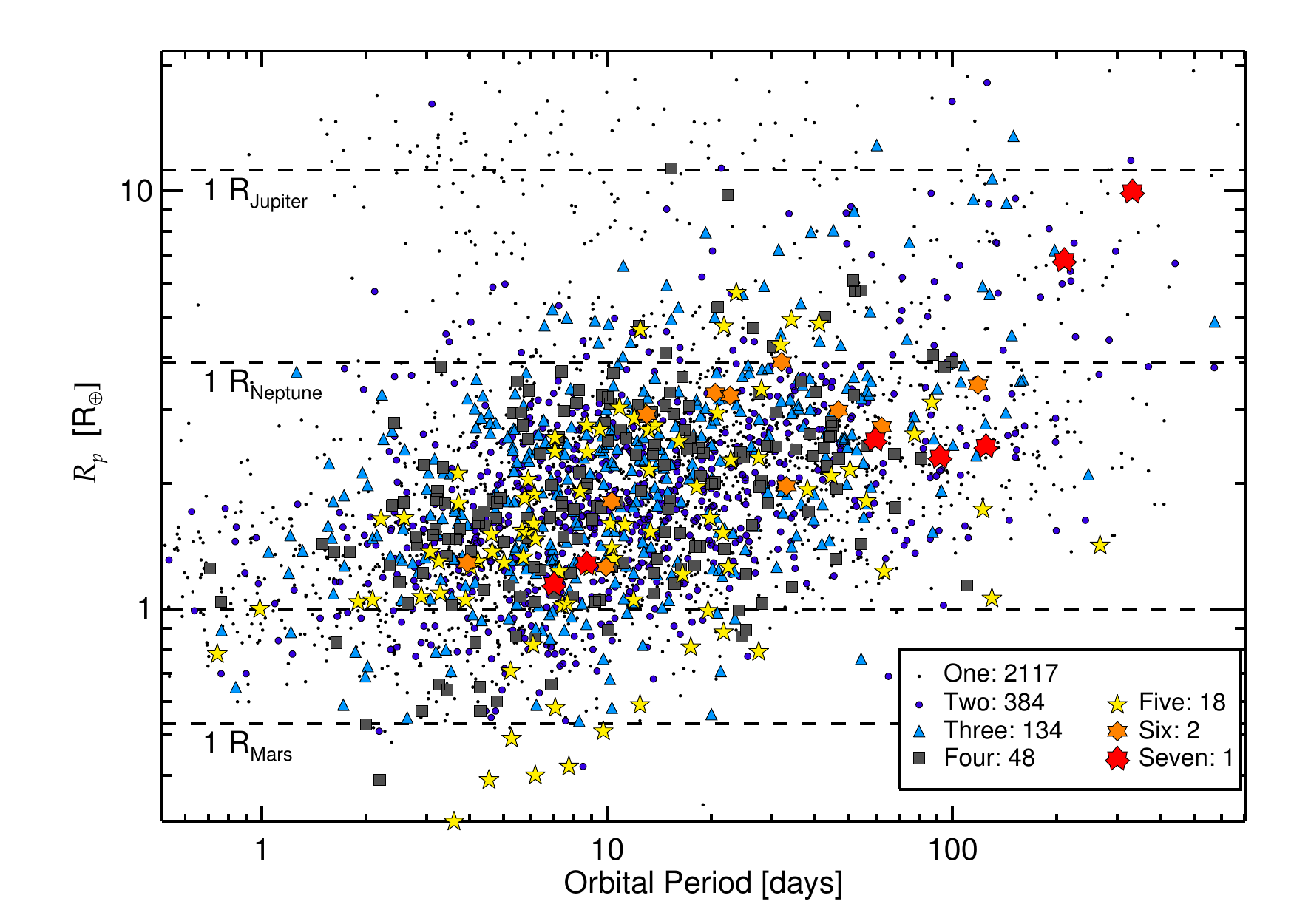}
\caption{
\label{fig:one}
\small
Orbital period versus planetary radius for planetary candidates found by analyzing the first three years of \ik data. Planets are shown by coloured symbols that represent the number of candidates in the system, as indicated by the legend in the lower right. The numbers shown in the legend represent the total number of systems of a given multiplicity in the catalog; a small fraction of these planets fall outside the boundary of the period-radius ranges plotted. Planets with shorter orbital periods are over-represented because geometric factors and frequent transits make them easier to detect in \ik data. The upward slope in the lower envelope of these points is caused by the difficulty in detecting small planets with long orbital periods, for which transits are shallow and few are observed. The apparent absence of giant planets in multi-planet systems has been quantitatively confirmed \citep{lat11}. Data provided by Jason Rowe.}
\end{center}
\end{figure}

Transits of $\sim3600$ planet candidates, the vast majority of which represent true exoplanets as described below, have been identified in the first three years of \ik data (Figure \ref{fig:one}). The discovery of these worlds, most of which have orbital periods (local ``years'') shorter than a few Earth months, has greatly expanded the zoo of known exoplanet types. Most \ik planets have radii, $R_p$, intermediate between those of Earth and Neptune (1 -- $3.8\rearth$, where $\rearth=6371$~km is the Earth's radius); planets in this size range are missing from our Solar System. These planets have a wide range of densities \citep{bat11, liss11a, doy11, cart12, jon14, mar14}, probably because they have atmospheres with a wide range of properties. Nonetheless, theoretical models of their interiors \citep[e.g.,][]{for07} imply that all of the planets in this class are ``gas-poor'', that is, less than half --- in most cases far less --- of their mass consists of hydrogen and helium (H/He). In contrast, H/He make up more than 98\% of our Sun's mass as well as substantial majorities of the masses of Jupiter, Saturn, and almost all known giant exoplanets with $R_p>9\rearth$.  

\ikt's primary mission is to conduct a statistical census of the abundance of planets as a function of planetary size, orbital period and stellar type. \ik has found that planets are common, with the number of planets in the extended solar neighborhood of our Galaxy being comparable to or larger than the number of stars \citep{dre13}. Of particular interest is $\eta_\oplus$, the average number of Earth-like planets per star. ``Earth-like'' means having a radius similar to that of Earth and receiving about as much energy flux from its host star as Earth receives from our Sun; see below for a more precise definition. With some extrapolation downward in size and longward in orbital period, \ik data suggest that $\eta_\oplus \sim 0.1$, although as discussed below, there is a broad range of estimates of this value. Previous studies \citep{fv05,sou11} have found giant planets to be much more common around stars that are richer in heavy elements relative to light gases; \ik data have shown that no comparable trend exists for small planets \citep{buc12,wang13}. Almost half of \ikt's planet candidates are in systems in which multiple transiting planets have been found. As discussed below, the large abundance of such systems implies that flat systems containing multiple planets on closely spaced orbits are quite common. This finding supports models of planet formation within a disc of material orbiting a star that date back to Kant and Laplace.

\ikt's prime mission ended in May 2013 with the failure of a second reaction wheel that made precise stable pointing away from the spacecraft's orbital plane impossible. Nevertheless, data analysis over the next few years is expected to reveal hundreds or even thousands of additional planet candidates, probably including some that extend the range of exoplanets to smaller sizes and longer periods (lower right in Figure 1), perhaps including true Earth analogs in size and period that orbit Sun-like stars. These additional planets, plus better estimates of planetary sizes and planet detectability, will allow for improved estimates of the population of planets within our Galaxy. Although the hobbled \ik spacecraft cannot observe its original star field any longer, it has been reprogrammed to continue its search for other worlds, with a focus on planets orbiting small stars with orbital periods of less than one month; \ik's new mission is dubbed ``K2''. Other space missions will expand and exploit \ikt's discoveries over the next decade. The European Space Agency (ESA) recently launched the Gaia astrometric spacecraft, which will determine precise distances to \ik planet-hosting stars, enabling more accurate determination of the sizes of these stars and their associated planets. NASA's Transiting Exoplanet Survey Satellite (TESS), scheduled to launch in 2017, will conduct an all-sky search for transiting planets around the nearest and brightest stars using small-aperture, wide-field optics \citep{rick14}. TESS planets will be easier to study with other space- and ground-based observatories than are \ik planets, most of which orbit much fainter stars. Searches for transiting planets in space will continue in the 2020's with ESA's PLATO mission, which will have an effective aperture almost as large as does \ikt, together with a much larger field of view. Analysis of data from these advanced observatories, together with associated theoretical studies, should advance the studies of exoplanets pioneered by \ik far beyond the mission's original goals. 

\section*{Transiting planets and eclipsing binary stars}

\noindent
The transit depth yields the ratio of the planetary radius to the stellar radius, and the repetition rate of transits tells us the planet's orbital period. The stellar colors --- or, better yet, stellar spectrum --- can be used to deduce the star's radius and mass, and from these we can find the planet's radius and the semi-major axis of its orbit (from Kepler's third law). In favourable cases (generally restricted to close-in planets that are subjected to intense stellar irradiation), we can detect the occultation of the planet as it travels behind the star, and thus determine the planet's albedo (i.e., its reflectivity). A wide range of albedos is found for both small \citep{dem14} and large planets \citep{ang14}, with most (hot) giant planets having low albedo. Planets in multiple systems perturb one another through their mutual gravity, causing their orbits to deviate from strict periodicity. These deviations lead to transit timing variations (TTVs) that in favourable cases can be used to measure the planetary masses and additional orbital elements \citep{hol10,liss11a,cart12,liss13,do14,jon14}. 

The objects in the catalogs assembled by the \ik project \citep{bor11,bat13,bur14} are considered to be only planet \emph{candidates} because eclipsing binary stars can mimic transiting planets. Normally the fractional brightness change in a binary-star eclipse is much larger than in a planetary transit, but occasionally the eclipse is grazing, or light from the \ik target star is diluted or ``blended'' with the light from a background or companion eclipsing binary that is nearby on the plane of the sky. Such false positives plague ground-based searches for exoplanets, but the \ik light curves (starlight received as a function of time) are of such high quality that they can usually be used to discriminate between grazing or blended stellar eclipses and planetary transits. Moreover, \ik is an imaging instrument, which can measure changes in the position of the image on the sky plane during transit. This ``centroiding'' weeds out most eclipsing binaries that are blended with background stars and some that are blended with companion stars. Thus, although well under half of \ikt's planet candidates have been verified to be true planets, the false-positive rate of the catalog as a whole is probably less than 10\%, although it may exceed 30\% for the largest planet candidates \citep{mj11,san12,fre13}. Therefore, with appropriate care, the \ik catalog can be used for statistical studies of the exoplanet population. 

Several dozen candidates not found by the \ik data analysis pipeline have been identified by eye by members of the public through the Planet Hunters project \citep{fis12}, and other groups have found dozens of planets with orbital periods of $< 1$ day for which the pipeline is not optimized \citep{san14}. 

Planet candidates that have been verified to be true planets at a high level of confidence (in most cases well above 99\%) are assigned \ik designations (names). Verification can take the form of dynamical confirmation by detection of either TTVs in the \ik light curve or radial-velocity variations, or it can be based on statistical arguments showing that the likelihood of the planet hypothesis is much greater than that of other possible causes of the observed light curve \citep{tor11,mor12,liss14}.

\section*{Individual planets and planetary systems}

\noindent
\ikt's primary mission is a statistical characterization of the exoplanet population, but we first describe some of the highlights of individual planets and planetary systems found by \ikt. 

\ikt's first major discovery was the Kepler-9 system, which contains two transiting giant planets with orbital periods of 19.24 and 38.91 days. The nearby 2:1 orbital resonance (38.91/19.24=2.02) induces TTVs of tens of minutes in both planets. Analysis of these TTVs enabled both planets to be confirmed and provided estimates of their masses: they are similar in size to Saturn but less than half as massive \citep{hol10,do14}.  TTVs have also been used to discover non-transiting planets, such as Kepler-19 c \citep{bal11} and Kepler-46 c \citep{nes13}. 

The first rocky planet found by \ik was Kepler-10~b, which has $R_p = 1.42\pm0.04\rearth$ and mass $M_p = 4.6\pm1.2\mearth$ \citep{bat11}, where $\mearth=5.978\times 10^{24}\,\mbox{kg}$ is the Earth's mass. This planet's density, $8.8^{+2.1}_{-2.9}\gcmc$, is consistent with an Earth-like composition. Remarkably, its orbital period is only 20 hours.

Kepler-11 is a Sun-like star with six transiting planets that range in size from $\sim 1.8$ -- $4.2\rearth$ \citep{liss11a,liss13}. Orbital periods of the inner five of these planets are between 10 and 47 days, with the ratio of orbital periods between adjacent planets ranging from 1.26 to 1.74. For comparison, the ratio of orbital periods in the Solar System ranges from 1.63 (Venus and Earth) to 6.3 (Mars and Jupiter). The outermost planet, Kepler-11 g, has a period of 118.4 days. TTVs have been used to estimate the planets' masses. Most if not all have a substantial fraction of their volume occupied by the light gases hydrogen (in the form of molecular hydrogen) and helium, which implies that H/He can dominate the volume of a planet that is only a few times as massive as the Earth.

Kepler-36 \citep{cart12} hosts two planets whose semi-major axes differ by only 10\% but whose compositions are dramatically different: rocky Kepler-36~b has a mass $ M_p = 4.5 \pm 0.3\mearth$, a density of 7.46$^{+0.74}_{-0.59}\gcmc$ and an orbital period, $P$, of 13.84 days, whereas puffy Kepler-36~c has $M_p = 8.7 \pm 0.5\mearth$, a density of 0.89$^{+0.07}_{-0.05}\gcmc$ and $P = 16.24$ days. Possibly the atmosphere of Kepler-36~b was stripped by photo-evaporation or impact erosion, while Kepler-36~c was able to retain its atmosphere because of its larger core mass and slightly larger distance from the host star \citep{lop13a}. The proximity of the orbits also presents a conundrum: while numerical integrations show that the current configuration may be long-lived, most nearby configurations are unstable on short timescales \citep{deck12}, so it is far from clear how these planets arrived at their current orbits.

The first transiting circumbinary planet to be discovered, Kepler-16 b, is an object of approximately Saturn's mass and radius ($M_p=106\pm5\mearth$, $R_p=8.27\pm0.03\rearth$), traveling on a nearly circular orbit (eccentricity $e=0.0069$) with a period of 228.8 days around an eclipsing pair of stars with an orbital period of 41.08 days \citep{doy11}. A bonus in such systems is that the planetary transits enable accurate measurements of the stellar masses and radii (errors $\lesssim 0.5\%$): one of the stars is about two-thirds the size and mass of our Sun and the other only a fifth as large as the Sun \citep{doy11}. Moreover, the primary star's rotation axis has been measured to be aligned with the binary's orbital axis to within 2.4$^\circ$ \citep{winn11}. Several other circumbinary planets have been found using \ik data, including the multi-planet Kepler-47 system \citep{oro12}.

Kepler-20 e is the first planet smaller than Earth ($R_p=0.87^{+0.08}_{-0.10}\rearth$) to be verified around a main-sequence star other than the Sun \citep{fre12}; its 6.1-day orbit means that it is far too hot to be habitable. The low-mass (M-dwarf) star Kepler-42 hosts three validated planets smaller than Earth, the smallest of which is Mars-sized \citep{mui12}. Kepler-37 b, only slightly larger than Earth's Moon, is the first planet smaller than Mercury to be found orbiting a main-sequence star; its period is 13 days, and the stellar host is 80\% as massive as the Sun \citep{bar13}. KIC 12557548 b exhibits transits of varying depths, which might be due to an evaporating dusty atmosphere \citep{rap12}. Kepler-78 b \citep{san13} has the shortest orbital period of any confirmed exoplanet, circling its star in 8.5 hours. This roasting world is slightly larger than Earth, and its mass, measured from the radial-velocity variations it induces in its nearby host star, implies a rocky composition \citep{how13,pepe13}.

Circumstellar habitable zones are conventionally defined to be the distances from stars where planets with an atmosphere similar to Earth's receive the right amount of stellar radiation to maintain reservoirs of liquid water on their surfaces \citep{kop13}. Kepler-62 f is the first known exoplanet whose size ($1.41\pm0.07\rearth$) and orbital position suggest that it could well be a rocky world with stable liquid water at its surface \citep{bor13}.

\section*{Principal goals of the \ik mission}

\noindent
Discoveries like the ones mentioned above have captured a great deal of attention in the scientific community and beyond. But \ik is, in essence, a statistical mission, designed to discover large numbers of planets in a survey with well-characterized selection criteria. The stated goals of the \ik mission prior to launch \citep{bor07} were to explore the structure and diversity of extrasolar planetary systems and thereby to: 

\begin{enumerate}

\item determine the frequency of Earth-size and larger planets in or near the habitable zone of a wide variety of spectral types of stars;

\item determine the distributions of size and orbital semi-major axes of these planets;

\item estimate the frequency of planets in multiple-star systems;

\item determine the distributions of semi-major axis, albedo, size, mass and density of short period giant planets;

\item identify additional members of each photometrically discovered planetary system using complementary techniques;

\item determine the properties of those stars that harbor planetary systems.

\end{enumerate}

In its four-year prime mission \ik observed a total of almost 200,000 stars, including $\sim140,000$ dwarf or main-sequence stars that were monitored for a substantial majority of the time. In addition to its contribution to exoplanet science, \ik has revolutionized the field of asteroseismology, which probes stellar interiors by observing the surface manifestations of oscillations that propagate within stars \citep{chap13,chr13}, and has dramatically advanced our understanding of eclipsing binary stars \citep{prsa11} as well as other areas of stellar physics. This article only considers stellar properties indirectly through their contributions to assessing planetary characteristics.

\section*{How common are planets?}
\label{sec:abundances}

\noindent
The \ik catalog of planets is uniquely valuable for studying the structure and properties of planetary systems: it is large enough that we can map out the distribution of planets in multiple parameters (orbital period, radius, multiplicity, properties of the host star, etc.); it has, at least in principle, well-defined selection criteria (in contrast to radial-velocity catalogs, which come from many different surveys and which usually do not include null results); and most of the parameter space that it explores --- typical radii of 1 -- $3\rearth$ and orbital periods up to $\sim 1$~year (Figure 1) --- is not easily accessible by other techniques. 

One of the most fundamental statistics describing planetary systems is the probability distribution $f(R_p,P)d\ln R_p\,d\ln P$ that a member of a specified class of stars possesses a planet in the infinitesimal area element $d\ln R_p\,d\ln P$ \citep{tab02,you11,how12,dong13}. The integral of this distribution over a range in planetary radius and orbital period is the average number of planets per star (not to be confused with the fraction of stars with planets, which is smaller). \ik determines this distribution for solar-type stars with reasonable accuracy for $R_p\gtrsim 1 \rearth$ in the range $P\lesssim 50$ days, and for $R_p\gtrsim 2\rearth$ in the range $P\lesssim 150$ days.

Occurrence rate calculations must carefully account for the completeness, reliability, and threshold criteria of the \ik catalogue \citep{jen10,ten12}, as well as random and systematic errors in the host-star properties. For candidates with small transit depths or just a few transits, robust estimates of occurrence rates require calibration by injecting and recovering planetary signals in \ik data \citep{jchr13,pet12,pet13b,pet13}. The false-positive frequency distribution must be modeled simultaneously \citep{fre13}. Revisions of host-star properties can dramatically alter the radius distribution of planets \citep{dre13}. No occurrence rate calculations to date contain all of these ingredients. Here we summarize key results from early analyses of the \ik dataset (see \citealt{bat14} for a review).

The studies cited above find that the number of planets per unit log period is nearly flat for $R_p\lesssim 4\rearth$ and $P > 10$ days, but rises by a factor 2 -- 5 between $P=10$ days and $P=100$ days for larger planetary radii. The number of planets drops sharply for orbital periods below 10 days for $R_p\lesssim 4\rearth$ and below 2 -- 3 days for giant planets. The occurrence rate of giant planets on small orbits is a factor of three lower than in radial-velocity surveys \citep{how12,wri12,fre13}, perhaps because a significant fraction of giant planets are injected into small orbits through planet-planet gravitational interactions, and the relatively metal-poor \ik stars host fewer and/or less massive planets, which are less likely to interact strongly \citep{daw13}. At all periods the number per log radius grows as the radius shrinks, at least down to radii of $2\rearth$. Below $2\rearth$, the distribution per unit log radius plateaus at orbital periods out to 50 days and probably out to 100 days \citep{dong13,pet13}. The average number of planets per star with $P<50$~days is $\approx 0.2$ for $1\rearth < R_p< 2\rearth$ and $\approx 0.4$ for all radii $R_p>1\rearth$ \citep{dong13,pet13b}. For stars cooler and less massive than the Sun, the average number of planets per star is even higher \citep{dre13}: $0.49^{+0.07}_{-0.05}$ for $1 \rearth < R_p< 2\rearth$ and $0.69^{+0.08}_{-0.06}$ for all radii $R_p>1\rearth$ with $P<50$~d. The higher frequency is remarkable since the fixed period cutoff at 50 d corresponds to a smaller semi-major axis in the less massive stars.

A widely used milestone is $\eta_\oplus$, the number of Earth-like planets of Sun-like stars. One difficulty in discussing $\eta_\oplus$ is that different authors use different definitions for ``Earth-like''. For solar-type stars the most natural definition is $\eta_\oplus=f(1\rearth,1\mbox{\,yr})$; for other stars we can replace $P=1\mbox{\,yr}$ with the period corresponding to the same incident stellar flux. Roughly speaking, this is the number of planets per star in a range of a factor of $e$ in radius and period centered on the Earth's radius and period. Unfortunately, determining $\eta_\oplus$ according to this definition requires an extrapolation downwards in size and longward in orbital period from the region where \ik has a statistically reliable planet sample, which introduces considerable uncertainty. Applying this extrapolation to power-law fits $f(R_p,P)\propto P^\beta$ \citep{dong13} of the distribution of planets in the 16-month \ik catalog \citep{bat13} yields $\eta_\oplus=0.09$. An independent analysis of \ik light curves \citep{pet13} gives a consistent result, $\eta_\oplus=0.12\pm0.04$, after renormalizing by a factor of 2.1 to convert their definition to ours. However, a follow-up study using the same catalogue but a more general form for the period and radius distribution found a much smaller value $\eta_\oplus=0.02^{+0.02}_{-0.01}$ \citep{dfm14}. There are also other uncertainties: for example, none of the results for $\eta_\oplus$ discussed here model false positives (for planets of this size, the biggest contributor is larger planets orbiting fainter stars that appear close to the \ik target on the plane of the sky; \citealt{fre13}). Moreover, the extrapolations involve a mix of planets with and without substantial volatile envelopes (see below), as well as a likely mix of formation histories, and therefore the extrapolation may not capture the true occurrence rate.

A related number for cool, low-mass stars is 0.16$^{+0.0.14}_{-0.10}$ planets per star with $0.5\rearth < R_p < 1.4 \rearth$ receiving 0.46--1 times the solar flux at Earth \citep{dre13}, which corresponds to a rate of $0.26^{+0.23}_{-0.16}$ in an interval equal to that we use to define $\eta_\oplus$. 

\section*{The diverse physical properties of \ik planets}

\label{sec:physicalprop}

\noindent
\ik has discovered more than 3000 planet candidates with radii $R_p<4\rearth$. Planetary interior models show warm planets of this size to be ``gas-poor," defined here as composed of less than 50\% H/He by mass. Transit surveys are well-suited to studying the physical properties of such planets because their radii are very sensitive to small amounts of gas in the atmosphere --- for example, just 1\% H/He added to a $1\mearth$, $1\rearth$ solid core can inflate the planet to $2\rearth$ \citep{lop13b} --- and moderately sensitive to the bulk composition of the core (e.g., water vs.\ rock). Furthermore, the gas-poor planets found by \ik are valuable because: (i) they sample a wide range of incident fluxes and therefore were presumably subject to a wide range of photo-evaporation rates; (ii) many are found at short orbital periods, allowing mass measurements or meaningful upper limits through radial-velocity follow-up studies \citep{mar14}, and (iii) some are found in compact systems with multiple transiting, low-density planets, whose short orbital periods and large radii allow sensitive mass measurements through TTVs. 

\begin{figure}[h!] \begin{center} \includegraphics[width=\columnwidth]{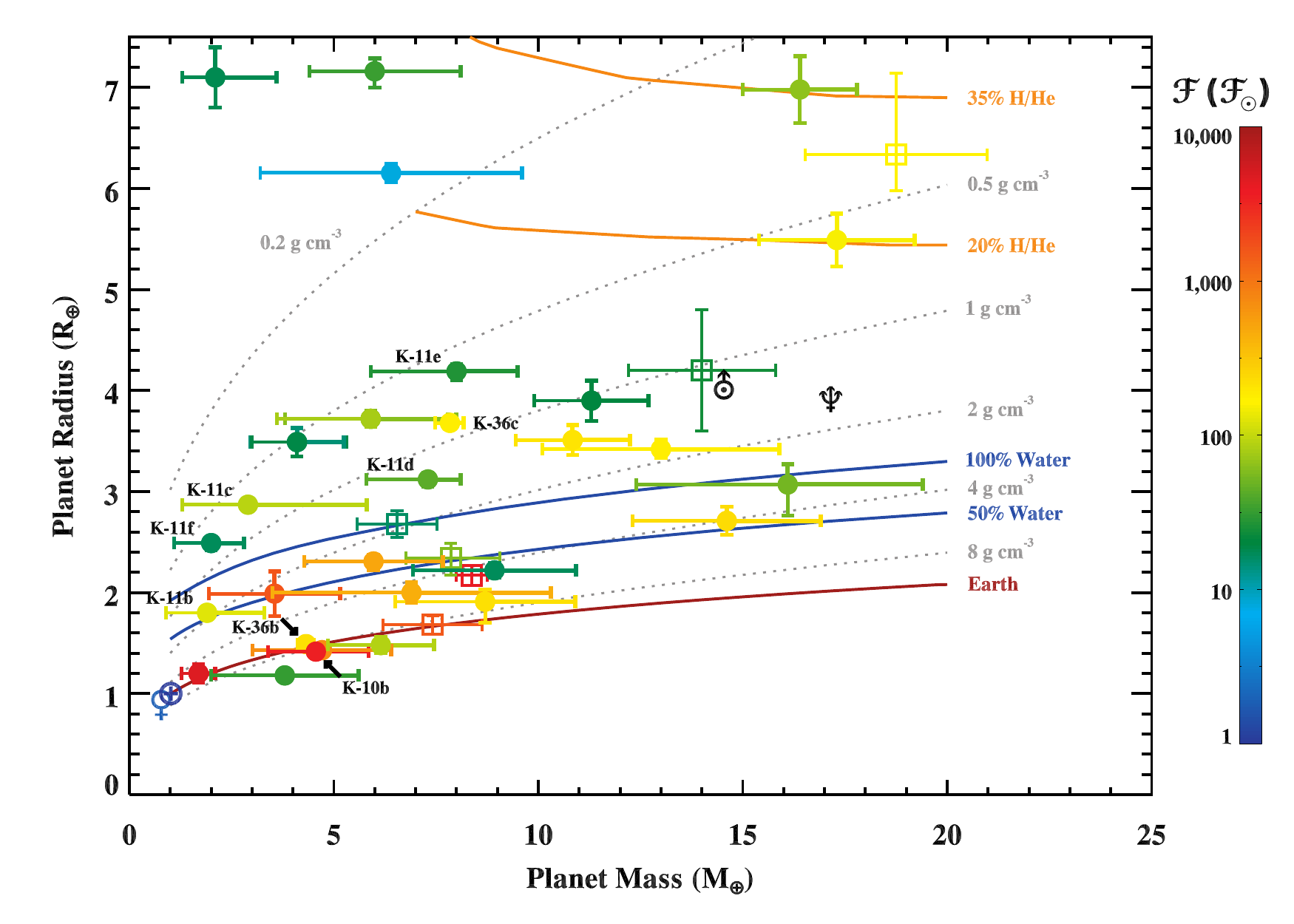}
 \caption{
\small
Mass-radius diagram for transiting exoplanets with measured masses less than $20\mearth$, along with model curves for different compositions. Planets are color-coded by the incident bolometric flux they receive. \ik planets are shown by filled circles, with numbers and letters indicating planets discussed in the text; the rocky planets in the crowded region near the lower left include Kepler-10 b  (red point) and Kepler-36 b  (yellow). Other known exoplanets in this mass range are shown by open squares. The Solar System planets Venus, Earth, Uranus and Neptune are shown by their symbols.  The lower curve is for an Earth-like composition with 2/3 rock and 1/3 iron by mass. All other curves use thermal evolution calculations \citep{lop12}, assuming a volatile atmosphere of H/He or water atop a core of rock and iron with composition the same as that of the bulk Earth. The two blue curves are for 50\% and 100\% water by mass and the two orange curves are for H/He atmospheres atop Earth-composition cores.  These theoretical curves assume a radiation flux 100 times as large as that received by Earth and an age of 5 Gyr.  Figure courtesy of Eric Lopez. 
\label{mrfig} 
}
\end{center}
\end{figure}

Figure \ref{mrfig} shows the masses, radii and incident flux received by well-characterized planets less than 20 times as massive as Earth. The wide range in size of gas-poor planets of a given mass indicates a diversity of composition. Note that most of the sub-Saturn exoplanets whose masses and radii are both known are \ik discoveries.

Various processes can affect the composition of gas-poor planets, including coagulation from volatile-rich or volatile-poor planetesimals, accretion of gas from the proto-planetary nebula if the planet forms before its dispersal, outgassing of volatiles from the planet's interior, atmospheric escape (e.g., via photo-evaporation), and erosion or enrichment of the atmosphere and mantle via collisions with planetesimals. Distinguishing which of these processes are at play and their relative contributions is both a challenge and a motivation for interpreting the measurements from \ikt.

Several of the planets discussed earlier and highlighted in Figure \ref{mrfig} have served as case studies to illuminate the properties of gas-poor planets. In particular, they sample a continuum of photo-evaporation rates, which are a function of both incident stellar flux (as a proxy for the X-ray/UV radiation responsible for atmospheric erosion) and core mass. The middle four planets of Kepler-11, each of which contains $\sim 4$ -- 15\% H/He by mass, might represent the pristine, uneroded initial compositions of gas-poor planets, whereas the innermost planet in the system, Kepler-11~b, is only 0.5\% H/He by mass (or maybe devoid of light gases entirely if it is water-rich), perhaps because it has undergone significant mass loss from its primordial atmosphere \citep{liss13}. Kepler-10~b, with an incident flux about 30 times that of Kepler-11~b, may in turn have lost all of its atmosphere to photo-evaporation. This speculation is based on Kepler-10~b's density, which can be matched by theoretical models that do not require a volatile component \citep{bat11}. In addition to lifetime-integrated X-ray and UV flux, core mass is an important factor in determining the photo-evaporation rate. A larger core mass for Kepler-36~c may have enabled it to maintain its atmosphere against photo-evaporation, which may have stripped its nearby neighbour Kepler-36~b \citep{lop13a}. 

There is now a large collection of gas-poor \ik planets with masses that are individually less precisely measured than the case studies above but nonetheless statistically powerful when analyzed as an ensemble. Dozens of masses have been measured via radial-velocity follow-up \citep{mar14}. Using an approach that accounts for degeneracies between mass and eccentricity \citep{lith12}, over 100 were estimated from TTVs \citep{hadd13}. Furthermore, theoretical models imply that radii of warm planets in the size range $2-4 \rearth$ depend far more on H/He percentage than on total planet mass \citep{lop13b}. Thus, the larger sample of thousands of \ik candidates with $R_p < 4\rearth$, even if lacking measured masses, informs us about planetary occurrence rates as a function of composition and their correlations with other properties such as the orbital period and the mass and chemical composition of the host star, which in turn constrain models for the formation and evolution of gas-poor planets. Several radius ranges may indicate different regimes for planet formation and evolution, as illustrated in Figure \ref{fig:comp}:

\begin{figure}[h!]
\begin{center}
\includegraphics[width=0.9\columnwidth]{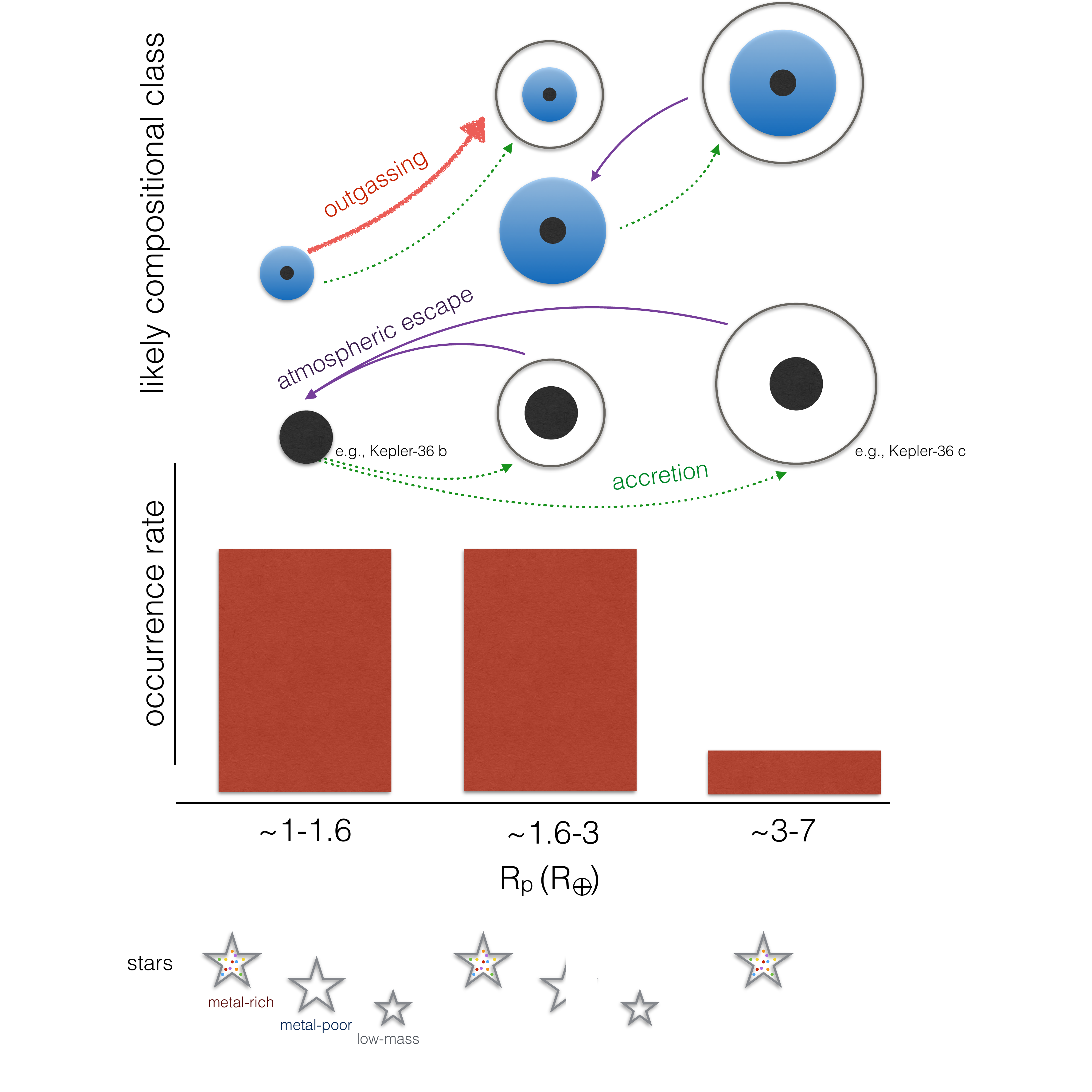}
\caption{\small Schematic illustration of plausible compositions of the small and mid-sized planets observed by \ikt, including rock (dark grey), water (blue), and light gases (white). The red bars indicate their approximate relative occurrence rates \citep{dong13,fre13,pet13}, and the arrows indicate physical processes that set or alter planet compositions. The smallest planets, on the left, can be rocky or possibly mixtures of rock and H$_2$O. Somewhat larger planets have volumetrically significant amounts of constituents less dense than rock. Planets whose sizes are comparable to or larger than that of Neptune, $R_p=3.8\rearth$, have envelopes composed of the lightest gases, H$_2$ and He. 
\label{fig:comp}}
\end{center}
\end{figure}

\begin{itemize}
\item $R_p\lesssim 1.6\rearth$: Most of the small number of  transiting planets in this radius range that have measured masses are dense enough to be rocky; in contrast, larger planets appear to require a volatile component \citep{rog14}. This radius range includes all verified \ik planets with period $P<1$ day \citep{san14}; these planets may once have possessed atmospheres that have now been stripped by impacts, photo-evaporation, stellar winds, and/or tidal forces.

\item $1.6\rearth \lesssim R_p\lesssim 3\rearth$: In this regime, the mass-radius relation is consistent with $M_p\propto R_p$, indicating that the typical planetary density decreases with increasing size \citep{wu13}. This mass-radius relation requires a substantial mass fraction of water or a small mass fraction (0.1 -- 5\%) in a H/He atmosphere \citep{wu13,wei14}. The scatter in the mass-radius relation exceeds the measurement errors \citep{wei14} indicating some diversity in composition and/or atmospheric properties, possibly including rare rocky planets without voluminous atmospheres. The presence of a H/He atmosphere substantially increases the temperature at the rocky surface; thus planets such as Kepler-22 b, which has a radius of $2.4\rearth$ \citep{bor12}, are unlikely to be habitable. 

\item $3\rearth \lesssim R_p \lesssim 7\rearth$: Planets in this size range are less dense than water, implying voluminous H/He atmospheres \citep{wu13}. The occurrence rate plummets between 2 and $3\rearth$ \citep{pet13,mor13}, so this class is much rarer than the first two. Few planets in this class have been found around low-mass stars \citep{wu13}. 

\item $R_p\gtrsim 4\rearth$: Large planets are more common around stars with larger abundances of elements heavier than helium \citep{buc12,wang13}.

\end{itemize}

The class of rocky planets with $R_p \lesssim 1.6\rearth$ may lack gaseous atmospheres either because they were insufficiently massive to accrete significant amounts of light gases in the regions of the protoplanetary disk where they formed and never outgassed an atmosphere, or because their primordial atmospheres were removed by giant impacts or photo-evaporation. A possible explanation for the plunge in occurrence rate between 2 and $3\rearth$ is that larger worlds need H/He envelopes, which are uncommon, but most low-density smaller planets contain substantial water components or (very low mass) outgassed H$_2$ envelopes. The paucity of planets larger than $3 \rearth$ around low-mass stars and the higher heavy-element abundance in the host stars of systems containing planets larger than $4 \rearth$ may both reflect the difficulty of accreting H/He envelopes in a protoplanetary disc with a low surface density of solids.

\section*{Properties of planetary systems}

\label{sec:systems} 

\noindent
The \ik catalog is particularly important because it contains many multiple-planet systems, roughly eight times as many as all radial-velocity surveys combined \citep{bur14}. Multiple systems are expected to have a very low false-positive rate ($\lesssim 1\%$), because background binary-star eclipses may mimic the light curve from a single transiting planet but are unlikely to imitate two or more \citep{liss12,liss14}. Multiple systems are also valuable because gravitational interactions among the planets lead to TTVs that in some cases allow us to determine the masses and orbital properties of one or more of the planets \citep{hol10, liss11a, cart12, liss13,do14,jon14}. 

Multiple systems also allow us to constrain the average mutual inclinations of the planetary orbits, either by comparing relative transit durations and orbital periods \citep{fab12} or by comparing the frequencies of systems with different multiplicities in the \ik survey to those in radial-velocity surveys \citep[since the chance that multiple planets in a single system will transit is much higher if their mutual inclination is low;][]{td12}. Such studies show that the typical mutual inclinations in \ik planets are only a few degrees, similar to the Solar System. Another probe is the mutual inclination between a transiting planet and its non-transiting perturber, which is not biased towards low mutual inclinations by the selection effects that are present in multi-transiting systems; the first such systems with good constraints have been found to be flat 
\citep{nes13,daw14}.

The finding that typical \ik multi-planet systems are flat is perhaps the first direct evidence that most planetary systems formed from a rotating thin disc of gas and dust, as suggested by Laplace over two centuries ago. But even this uncontroversial result leads to tension with other observations. In most formation models, planets have mean inclinations that are at least half as large as the mean eccentricities \citep[e.g.,][]{ida90}, and this result also holds for the planets in the Solar System, the asteroids, and the Kuiper belt. Thus we expect the mean eccentricity of the \ik planets to be no more than about 0.1. Unfortunately, attempts to measure the eccentricity distribution of \ik planets have been complicated by (or have sometimes brought to light) systematic uncertainties in the stellar properties \citep{moor11,pla14,kip14b}, though individual constraints have been possible for a subset of well-characterized stars with high signal-to-noise transits \citep{kip12,dj12}. In contrast, the eccentricities of radial-velocity planets are straightforward to measure, and the mean eccentricity for those having orbital periods larger than 10 days is 0.26, far larger than we would expect from the arguments above. Are the eccentricities and inclinations of the radial-velocity planets larger than those of the \ik planets? Or perhaps just of those planets in \ikt's multiple planet systems? And if so, why? Might the eccentricities be over-estimated \citep{zak11}?  Or could exoplanets have much larger eccentricities than inclinations \citep{raf10}?

An equally serious tension is revealed by ground-based measurements of the stellar obliquity, the angle between the equator of the host star and the orbital plane of a transiting planet. About 80 obliquities --- or at least their projections on the sky plane --- have been determined, mostly through measurements of the Rossiter--McLaughlin effect \citep{alb12}. Almost half of the measured projected obliquities exceed $20^\circ$ and 15\% exceed $90^\circ$; of course the width of the distribution of true (as opposed to projected) obliquities must be even larger. This result is quite different from the expectation for a Laplace-type model, in which the host star and planets form from a single rotating gas disc and thus should have a common spin and orbital axis. One possibility is that the close-in giant planets arrived on their present orbits via high-eccentricity migration, which excites large obliquities \citep{ft07}. Another possibility is that the stellar spin is misaligned with the axis of the planetary disc, perhaps because of a collision with a giant planet on a highly eccentric orbit or twisting of the planetary disc by external torques \citep{tre91,hel93,bat12}. To complicate the situation further, most of the handful of \ik planets for which measurements are available, including multi-planet systems, have obliquities near zero \citep{sj12,chap13,hir14}.

\begin{figure}[h!]
\begin{center}
\includegraphics[width=.7\columnwidth]{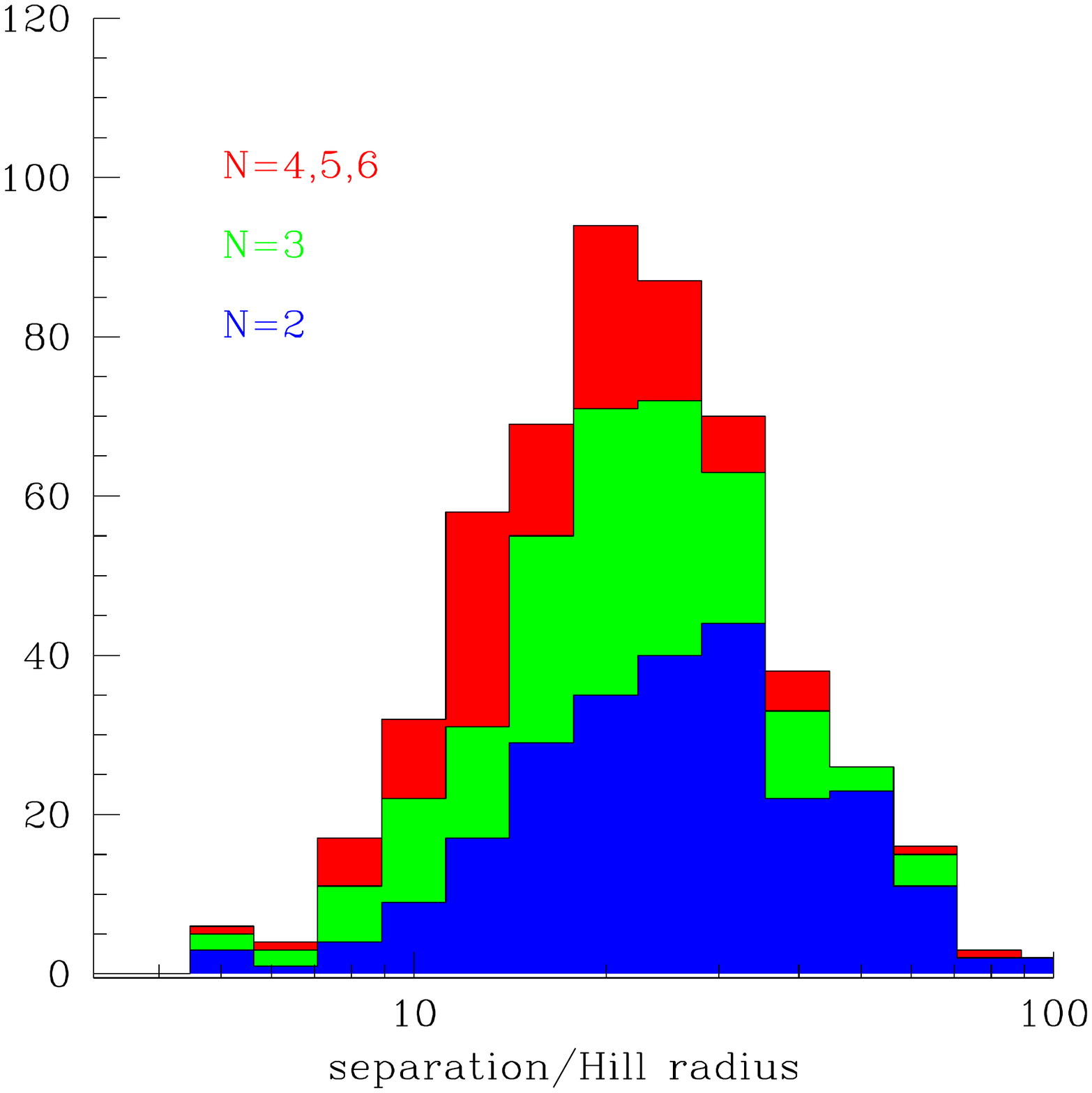}
\vspace{-2.0cm}
\caption{ \small Separations of nearest neighbours in the \ik multi-planet systems, measured in Hill radii (eq.\ \ref{eq:hill}), from the catalogue of \cite{fab12}. Masses are derived using the mass-radius relation $M_p=\mearth(R/\rearth)^\alpha$ with $\alpha=3$ for $R_p<\rearth$ and $\alpha=2.06$ for $R_p>\rearth$. Planets on circular, coplanar orbits separated by more than 9 -- 12 Hill radii are expected to be stable for the lifetime of typical stars; the few planet candidates seen with smaller separations may have large errors in their estimated radii, may not obey the assumed mass-radius relation, or may not be planets orbiting the same star. 
\label{fig:hill}
}
\end{center}
\end{figure}

The properties of multi-planet systems are constrained by the requirement that they be dynamically stable over timescales comparable to the lifetime of the star. Rigorous stability criteria are not usually available except for two-planet systems \citep{glad93}, but a useful approximate criterion is that systems composed of planets on nearly circular, nearly coplanar orbits are stable over $N\gg 1$ orbits if the separation in semi-major axis between adjacent planets, $a_{i+1}-a_i$, exceeds some constant $K_N$ times the mutual Hill radius (the separation at which the mutual planetary gravity equals the difference in the pull of the star on the two planets),
\begin{equation}
R_{H,i,i+1}\equiv\left(\frac{M_i+M_{i+1}}{3M_\star}\right)^{1/3}\frac{a_i+a_{i+1}}{2} ,
\label{eq:hill}
\end{equation}
where $M_{i,i+1}$ are the planet masses and $M_\star$ is the mass of the host star. For $N=10^{10}$, $K_N \approx 9$ -- 12 \citep{sl09}. As one would expect, most of the \ik multi-planet systems are safely stable by this criterion (see Figure \ref{fig:hill}), and numerical integrations assuming initially circular, coplanar orbits confirm that virtually all of them are stable for at least $10^{10}$ orbits \citep{liss11,fab12}. These results depend on the assumed mass-radius relation, but are relatively insensitive to it since the planet mass enters the definition of the Hill radius to the $1/3$ power.

A deeper question is whether these systems are dynamically ``full" or ``packed", which we define to mean that no additional planets, even with very small masses, could be inserted between the existing ones in a stable orbital configuration. The situation in our own Solar System is ambiguous: the region from Jupiter to Neptune is packed, or nearly so \citep{hol97}, but inside Mercury and between Earth and Mars there are significant bands in semi-major axis where additional low-mass planets would be stable for at least $10^8$ years \citep{et02}. 

Dynamically packed systems are a natural consequence of models in which planets grow hierarchically, since planet growth should stop once all of the orbits are stable. However, the correspondence is not exact, since stable zones may not be occupied if the planets they contained collided in the final stages of hierarchical growth; moreover, the system may contain additional planets that are not transiting or fall below the \ik detection threshold. The evidence from Figure \ref{fig:hill} suggests that the known \ik planets are typically separated by about twice the distance required for stability, but given the possible presence of undiscovered planets and the destabilizing effects of non-zero eccentricity, many of these systems may be dynamically packed \citep{fm13}.

Planets are also found in binary star systems, either orbiting one of the two stars with an orbital period much shorter than the binary period (``\emph{S-type}'') or orbiting both with a period much longer than the binary period (``\emph{P-type}'' or ``\emph{circumbinary}''). Most of our understanding of S-type planets comes from radial-velocity studies, whereas circumbinary planets around normal stars were first discovered by \ikt. In most respects the properties of planetary systems in single and binary-star systems are similar \citep{egg10,rag10}, although S-type planets appear to be less common in binary systems  than similar planets around single stars \citep{wang2014}. 

Binary systems offer unique insights into the formation of both planets and stars. (i) Torques from the companion star in an S-type binary can excite slow, large-amplitude Lidov-Kozai oscillations in the eccentricity and inclination of the planetary orbit. One striking hint that Lidov-Kozai oscillations are sometimes at work is that the four planets with the largest eccentricities ($e>0.85$) are all members of wide S-type binary systems \citep{tam08}. (ii) A close companion star truncates the protoplanetary disc and the planetary system at a radius of about 0.25 -- 0.3 times the companion's separation (for equal-mass stars on a circular orbit, \citealt{hw99}). No S-type planets have been discovered, either in radial-velocity or transit surveys, in binary systems with separation less than about 10 times the Earth-Sun distance. This could mean that the site of planet formation in the protoplanetary disc is beyond $0.3\times 10=3$ times the Earth-Sun distance, consistent with theories in which planets found at smaller radii have migrated inward; alternatively, the outermost regions at which circumstellar orbits are stable may nonetheless be too perturbed for planets to form. (iii) If binary stars form through dynamical interactions between single stars in a dense gas-free cluster, it would be difficult for them to acquire circumbinary planets. On the other hand, if binaries form through fragmentation and collapse in a gas-rich environment, they are likely to acquire a circumbinary disc in which planets could form. (iv) Binary stars with orbital periods of a few days are likely to be formed from binaries with much longer periods through high-eccentricity migration induced by a tertiary companion \citep{ft07}. This process would probably remove or destroy any planets initially orbiting one of the two stars and is unlikely to produce circumbinary planets detectable by \ik; therefore we should not expect to find planets, S-type or P-type, in binary systems with periods of a few days or less, and this expectation is so far confirmed by the observations --- the shortest-period planet-hosting binary star is Kepler-47, with a period of 7.45 d \citep{oro12}.

Just as important as the discoveries made by \ik are its non-discoveries. So far \ik has found no co-orbital planets, planets sharing the same average semi-major axis like the Trojan asteroids found accompanying Jupiter and the Saturnian satellites Janus and Epimetheus. It has found neither exomoons nor ``binary" planets orbiting one another \citep{kb12,kip13}. 

\section*{Planet formation}

\noindent
The mass fraction of stellar material other than the dominant elements of hydrogen and helium (the metallicity, in astronomical parlance) ranges from a few percent down to $\sim 0.01\%$ among stars in the solar neighbourhood. The initial protostellar disc presumably has the same composition as its host star, and these heavier elements are the raw material from which most of the mass in a typical \ik planet must be drawn. Thus it is natural to expect that planet formation should be easier around stars having high metallicity. This expectation is confirmed for the giant planets detected in radial-velocity surveys: the fraction of high-metallicity stars hosting giant planets is much larger than the fraction of low-metallicity stars \citep{san01,san04,fv05,sou11}. Similarly, a star in the \ik catalog with super-solar metallicity is $\sim 2.5$ times more likely to host a large planet ($R_p>5\rearth$) than a star with sub-solar metallicity \citep{wang13}. Remarkably, there is no such correlation for small planets ($R_p<2\rearth$): the probabilities that a \ik star with super-solar or sub-solar metallicity hosts a small planet are approximately equal \citep{wang13}, with a significant number of small \ik planets found around stars with metallicity as small as one-quarter that of the Sun \citep{buc12}. Perhaps this is a hint that the formation process for small planets has more than enough metals to draw on even in moderately low metallicity discs. Before any firm conclusions are drawn, we need reliable metallicities for a larger sample of \ik stars and planet-frequency measurements for stars with a wider range of metallicities. 

One of the most basic questions about the \ik planets is whether they formed in situ \citep{chi13,hm13} or migrated to their current orbits from larger radii \citep{rog11,swi13}. The orbital periods of most of \ikt's planets are $\lesssim 50$ days, resulting in nominal formation timescales that are much shorter than the lifetime of the gas disc, unlike those of the Solar System's terrestrial planets. Thus the planets and their gaseous envelopes could have formed in situ. The main argument for migration is that it is a robust process \citep{gt80} that inevitably occurs in both analytic models and numerical simulations of planets orbiting in gaseous discs \citep{baru13}. On the other hand, models of migration have not successfully predicted any populations of planets before they were observed. 

Additional insight into the migration process comes from planets in mean-motion resonances. In the strongest of these, the orbital periods of the two resonant planets are in the ratio $(n+1):n$, where $n$ is an integer. Planets can be captured permanently into resonance if they cross the resonance during migration and the migration is convergent, i.e., in a direction such that the period ratio evolves towards unity, rather than away. Capture into resonance during convergent migration is certain if the migration is slow enough and the planetary eccentricities are small enough \citep{pea76}. It is therefore striking that the multi-planet systems discovered by \ik contain very few resonant planet pairs: the excess fraction of planet pairs in the \ik sample having period ratios within 5 -- 10\% of 3:2 or 2:1 is less than 5\%. Possible explanations for the small fraction of resonant planets include the following: (i) migration was too fast for capture to occur; however, this requires migration times of $\lesssim 10^3$ years, far shorter than is plausible with disc migration \citep{rein12}; (ii) stochastic torques on the migrating planet, which might arise in a turbulent protoplanetary disc \citep{rein12}, allowed the planets to escape the resonances and continue migrating; (iii) eccentricity damping from the protoplanetary disc led to escape from the resonance \citep{gs14}; this mechanism requires that the \ik planets have very small eccentricities, since eccentricities are difficult to excite after migration is complete \citep{pet14}; (iv) perhaps the planets formed in situ rather than migrating. A possible clue to the answer, still poorly understood, is that the distribution of period ratios in the \ik multi-planet systems is asymmetric around the strong 2:1 and 3:2 resonances, with a peak just outside the resonance and/or a valley inside \citep{fab12,lw12,pm13}. 

\section*{Unsolved problems}

\noindent
\ik represents a watershed in our understanding of extrasolar planets and a great stride forward in understanding the properties of planetary systems and the problems in developing theories of their formation. Yet there are many aspects of planetary systems that \ik has not illuminated at all. \ik has opened up a new region in the orbital period vs.\ radius plane, containing planets as small as Earth's Moon at short periods, and larger planets with orbital periods as large as 1 -- 2 years, but \emph{all} of the planets in the Solar System lie outside this region (though only just outside in the case of Venus and Earth). The atmospheres of giant planets must be investigated through transit observations by ground-based telescopes or the Hubble and Spitzer space telescopes, as \ik has no spectral resolution. The eccentricities of planetary orbits provide important insights into their formation, but only ground-based radial-velocity surveys can routinely measure eccentricities. These are expensive because the \ik host stars are so faint, and often impossible with current technology because of the small masses of typical \ik planets. \ik cannot detect multiple transits of planets with periods longer than a few years, so the region beyond a few times the Earth--Sun distance, where most giant planets are likely to be born, is no better understood now than before. Planets are found occasionally at much larger radii and many have probably been ejected into interstellar space, and these regions can only be investigated by high-resolution imaging and gravitational microlensing.

\ik has hugely advanced our understanding of the phenomenology of exoplanets, but so far has led us no closer to a secure theory of planet formation. Did the \ik planets form in situ  or did they migrate from larger radii? Why are small planets common around host stars with such a wide range of metallicities? How did the \ik planets acquire their voluminous atmospheres, and why are the atmospheres so diverse in mass fraction and composition? How are the \ik planets related to the terrestrial planets in the Solar System? Why does the typical inclination of the (small) \ik planets appear to be much less than the typical eccentricity of the (large) radial-velocity planets? How are the large angles between some planetary orbital planes and the host-star equators generated? What is the relationship between the dynamics and formation of small, rocky planets and gas-giant planets \citep{sch14}? 

\section*{After \ik}

\noindent
Although data acquisition by the \ik spacecraft on its original target field has ended, ongoing data analysis and observational follow-up will refine the results obtained already and address some of the outstanding questions reviewed here.

Better models of the stellar and instrumental noise in the transit light curves, including a rigorous treatment of temporally correlated noise, may enable the discovery of smaller, longer period planets \citep{smi12,stu12} as well as better characterization of existing ones. More accurate transit times are crucial given that the majority of \ik planet masses are derived from TTVs, and more secure detections of and upper limits on transit duration variations (TDVs) will provide important constraints on mutual inclinations.

We can hope to address some of the unanswered questions about occurrence rates and system architectures by more sophisticated statistical analyses. The most complete view of the architectures of \ik planetary systems will require tying together constraints from occurrence rates, transit durations, TTVs, and TDVs. Even the absence of TTVs can provide important constraints on planetary system architectures --- for example, close-in giant planets appear to have fewer (or no) companion planets compared to more distant or smaller planets \citep{ste12}. Insight can be gained into the composition of gas-poor planets by joint modeling in the space of radius, incident flux, and (where available) mass (Figure \ref{mrfig}). 

Observational follow-up of \ik targets from the ground and space is underway. High-quality spectra of \ik host stars will help refine estimates of their properties, and therefore reduce the uncertainty in stellar properties that often dominates the uncertainty in planetary properties. A spectrum can also determine the star{'}s projected rotational velocity which, combined with the stellar spin period from \ik photometry, constrains the angle between the stellar equator and planetary orbit \citep{hir14}. Spectra and adaptive-optics imaging will allow the catalogs to be culled more completely of false positives. The determination of accurate host-star metallicities, which already has provided new insights into the planet-formation process \citep{buc12,wang13}, will be greatly expanded by LAMOST \citep{dong13a}. Programs are being developed to follow up some \ik candidates with large TTVs and establish a longer baseline for TDVs using ground-based telescopes.
\ik stars are among the billions astrometrically monitored by the Gaia mission, launched by ESA in December 2013. Gaia will determine the distances of the \ik target stars, thereby improving our knowledge of stellar parameters and, consequently, the planetary radii; better radii will in turn improve estimates of the planetary occurrence rates and compositions, and correlations between planetary and stellar properties. For stars within $\sim 200$ pc, both within and outside the \ik field, Gaia can detect Jupiter analogues by the astrometric oscillations of their host stars, revealing a more complete architecture for systems in which only the close-in planets are detectable by transits or radial-velocity measurements. 

Space-based all-sky surveys, including NASA's Transiting Exoplanet Survey Satellite (TESS; scheduled to launch in 2017) and ESA's PLAnetary Transits and Oscillations of stars (PLATO, scheduled to launch by 2024) could greatly increase the science returns from \ik by revisiting the \ik field. Such follow-up would provide a long time baseline that would allow for improved occurrence rates, including an accurate value for $\eta_\oplus$, masses for longer period planets from TTVs that would help address outstanding issues about their formation and composition, and the possibility of a substantial number of TDV detections. 

Two spacecraft scheduled to be launched in this decade will study known transiting planets. ESA's CHaracterising ExOPlanets Satellite (CHEOPS) will target known exoplanet hosts to discover additional transiting gas-poor planets. NASA's James Webb Space Telescope (JWST) will characterize the atmospheres of gas-poor planets analogous to those that \ik discovered in abundance; this may break the degeneracy among several composition possibilities. 

\bigskip

Since the first handful of exoplanet discoveries two decades ago, the pace of exoplanet research has been extraordinary, driven primarily by ground-based radial-velocity searches of growing power and sophistication and by NASA's \ik mission. The armada of projects described above will probe new regions of exoplanet parameter space and provide more detailed and accurate probes of the properties of known exoplanets. The challenges for the next two decades will be to maintain the momentum that has been built up in the last two decades of exoplanet research, and to work towards the even longer term goal of producing an image of an extrasolar planet comparable to the iconic images of Earth taken by the Apollo astronauts. 

Acknowledgments: This research has made use of the Exoplanet Orbit Database at exo\-planets.org \citep{wri11} and the Extrasolar Planets Encyclopedia at exoplanets.eu. We are grateful to the \ik Science Team for their extensive efforts in producing the high-quality dataset that has made possible the results reviewed here.  We thank William Borucki, Eugene Chiang, Subo Dong, Eve Lee, Eric Lopez, Leslie Rogers, Jason Rowe, Andrew Youdin and Kevin Zahnle for helpful discussions and comments on the manuscript. R.I.D. and S.T. gratefully acknowledge funding from the Miller Institute for Basic Research in Science at the University of California, Berkeley. 

\bibliography{converted_to_latex_v5}

\end{document}